\documentclass[11pt]{article}
\usepackage[abs]{overpic}
\usepackage{amsmath,amssymb,amsfonts}
\usepackage{graphicx,color}
\usepackage{setspace} 
\usepackage{hhline}

\makeatletter

\@addtoreset{equation}{section}
\makeatother

\newcommand{\be}{\begin{equation}}

\newcommand{\ee}{\end{equation}}
\newcommand{\bea}{\begin{eqnarray}}
\newcommand{\eea}{\end{eqnarray}}
\newcommand{\beann}{\begin{eqnarray*}}
\newcommand{\eeann}{\end{eqnarray*}}

\newcommand{\ba}{\begin{array}}
\newcommand{\ea}{\end{array}}

\newcommand{\N}{{\cal N}}
 
\def\XXint#1#2#3{{\setbox0=\hbox{$#1{#2#3}{\int}$} 
\vcenter{\hbox{$#2#3$}}\kern-.5\wd0}} 
 
\renewcommand{\thefootnote}{\fnsymbol{footnote}}

\begin{document}

\setlength{\oddsidemargin}{0cm}
\setlength{\baselineskip}{7mm}
\vfil\eject
\setcounter{footnote}{0}
\begin{flushright}
\normalsize
RIKEN-TH-181\\
\end{flushright}
    \begin{Large}  
       \begin{center}
         { \bf Triality in $SU(2)$ Seiberg-Witten theory \\
and \\
Gauss hypergeometric function
}
       \end{center}
        \end{Large}
        
\begin{center}
Ta-Sheng Tai\footnote[1]{
        e-mail address: tasheng@riken.jp
    }

\end{center}
\renewcommand{\thefootnote}{\arabic{footnote}}
\begin{small}
\begin{center}
{\it Theoretical Physics Laboratory, RIKEN,
                    Wako, Saitama 351-0198, JAPAN}
\end{center}
 \end{small}
\begin{abstract}
\noindent
{\normalsize 
Through AGT conjecture, 
we show how triality observed in $\N=2$ $SU(2)$ $N_f=4$ QCD can be interpreted geometrically as 
the interplay among six of Kummer's twenty-four 
solutions belonging to one fixed Riemann 
scheme in the context of 
hypergeometric differential equations. 
We also stress that our 
presentation is different from the usual crossing 
symmetry of Liouville 
conformal blocks, which is described by the 
connection coefficient in the case 
of hypergeometric functions. Besides, upon 
solving hypergeometric differential 
equations at the zeroth order 
by means of the WKB method, a curve 
(thrice-punctured Riemann sphere) emerges. The 
permutation between these six 
Kummer's solutions then boils down to the outer automorphism 
of the associated curve.
}
\end{abstract}
\setstretch{1.1}
\section{Introduction}
Adding 
$N_f$ massless fundamental hypermultiplets (flavors) to 
pure $\N=2$ $SU(2)$ Yang-Mills theory results in 
$SO(2N_f)$ 
flavor symmetry which gets 
enhanced to $Spin(2N_f)$ at the quantum level. This is 
due to the fact that 
monopoles in the low-energy 
Coulomb phase transform as spinors of 
$Spin(2N_f)$ once 
$2N_f$ fermionic zero-modes 
(collective coordinate) on them are quantized. 
In particular, when $2N_c=N_f=4$ where 
both vanishing one-loop $\beta$-function 
and exact scale invariance follow one sees that 
the outer automorphism group 
${\mathbf S}_3$ of $Spin(8)$, a double-cover of $SO(8)$, 
gets realized as a kind of $S$-duality, i.e. $triality$. 
Namely, its action in 
Coulomb phase permutes three fundamental BPS objects 
corresponding to three eight-dimensional 
irreducible representations of $SO(8)$, 
i.e. $(\mathbf{v,s,c})\equiv 
{\text {(electron, monopole, dyon)}}$. 
Our aim in this short letter is to clarify its geometric 
origin in terms of Gauss hypergeometric functions.

In \cite{Seiberg:1994aj} 
Seiberg and Witten wrote down explicitly 
the $SU(2)$ $N_f=4$ Seiberg-Witten (SW) curve $\Sigma$ 
parameterized by 
four bare flavor masses and 
$\tau_0 \equiv\dfrac{\theta}{\pi} + 
\dfrac{8\pi i}{g^2}$ (marginal bare gauge coupling)%
\footnote{ In the presence of 
fundamental flavors, $\tau_0$'s normalization 
deviates from 
the pure $\N=2$ one by a factor 2.}. 
They discovered the following exotic transformation rule 
attributable to the property of triality: 
\begin{eqnarray} 
S: ~\tau_0 \to -\frac{1}{\tau_0}, ~~~~~~
\begin{cases}
m_1 \to \frac{1}{2}(m_1 + m_2 + m_3 - m_4)\\
m_2 \to \frac{1}{2}(m_1 + m_2 - m_3 + m_4)\\
m_3 \to \frac{1}{2}(m_1 - m_2 + m_3 - m_4)\\
m_4 \to \frac{1}{2}(-m_1 + m_2 + m_3 - m_4)
\label{ma1}
\end{cases}
\end{eqnarray}
\begin{eqnarray} 
T: ~\tau_0 \to {\tau_0 +1}, ~~~~~~
\begin{cases}
m_1 \to m_1\\
m_2 \to m_2\\
m_3 \to m_3\\
m_4 \to -m_4
\label{ma}
\end{cases}
\end{eqnarray}
under which $\Sigma$ is kept invariant. 
This is because 
combinations of $S$ and $T$ together 
generate $SL(2,{\mathbb Z})/\Gamma(2)=\{ I, ~S, ~T, 
~ST, ~TS, ~STS \}$ which is identical to 
the outer automorphism group ${\mathbf S}_3$. Notice that 
these rules arising from observation still lack 
rigorous derivation. It is then seen that 
full $SL(2,{\mathbb Z})$ invariance w.r.t. $\tau_0$ 
shrinks to a smaller $\Gamma(2)$ one by including flavor 
mass deformations unless 
\eqref{ma1} and 
\eqref{ma} are taken into account. 
Explaining triality in a more 
geometric way has been attempted ever since 
$\N=2$ $SU(2)$ low-energy Coulomb phase 
dynamics got rephrased in, say, 
Vafa's F-theory setup 
\cite{Vafa:1996xn, Sen:1996vd} or Gaiotto's picture. 
Either one seems promising 
because totally geometric reformulations of a class of $\N=2$ 
theories were provided there.

The latter due to Gaiotto's seminal paper \cite{Gaiotto:2009we} 
interpreted nicely $S$-duality group of a large family of 
$\N=2$ superconformal quiver gauge theories 
as physically equivalent ways of deforming certain genus-$g$ 
$n$-punctured Riemann surface $C_{g,n}$%
\footnote{Though at first sight $C_{g,n}$ seems an 
ultra-violet object, a $r$-sheeted cover of 
it turns out to be the infra-red Seiberg-Witten curve 
of $A_{r\text{-}1}$-type $SU(r)$ SCFTs. }. Also, $C_{g,n}$'s 
moduli (Teichm{\"u}ller) space is accordingly identified with 
the space of a set of ultra-violet coupling constants, say, 
$q_{UV }=e^{\pi i\tau_{UV}}$. 
Based on his idea, 
the previous ${\mathbf S}_3$ has to be 
thought of as 
${\mathbf S}_4/({\mathbb Z}_2 
\times {\mathbb Z}_2)$ 
(${\mathbb Z}_2 
\times {\mathbb Z}_2$: Klein four-group or Vierergruppe) 
which manages to permute marked punctures on 
$C_{0,4}$ responsible for the $2N_c=N_f=4$ case. 
More precisely, from two $SO(4)$ ($SO(4)\times SO(4)\sim SU(2)_a \times SU(2)_b \times 
SU(2)_c \times SU(2)_d$) of $SO(8)$ one can 
decompose $\mathbf{8}$ as 
\begin{eqnarray*}
\mathbf{8}\sim (\mathbf{2}_a\otimes \mathbf{2}_b) 
\oplus (\mathbf{2}_c\otimes \mathbf{2}_d);
\end{eqnarray*}
therefore the action of 
triality exchanging three $\mathbf{8}$'s 
results in permuting punctures 
labeled by $SU(2)_\xi$ 
$(\xi=a,b,c,d)$ respectively. 
As pointed out later by Alday, Gaiotto 
and Tachikawa \cite{Alday:2009aq}%
\footnote{See also 
\cite{Alday:2010vg}-\cite{Nanopoulos:2009xe} for recent developments along AGT conjecture.}, further associating 
every puncture with a mass parameter 
$\mu$ subject to 
\begin{eqnarray}
&&m_1=\mu_a +\mu_d - \frac{Q}{2},~~~~~~~
m_2=-\mu_a+\mu_d +\frac{Q}{2},\nonumber\\
&&m_3=\mu_c +\mu_b -\frac{Q}{2},
~~~~~~~m_4=-\mu_c+\mu_b +\frac{Q}{2},
\label{mu0}
\end{eqnarray}
one easily agrees 
that \eqref{ma1} and \eqref{ma} can be completely accounted for 
by permutations of $\mu$'s with $Q=0$. Henceforth, 
\eqref{mu0} nowadays referred to as 
``AGT dictionary" opens up 
a new perspective for understanding triality geometrically%
\footnote{See \cite{Eguchi} for another geometric interpretation of triality resulting 
from E-string formalism.}. 
In fact, by introducing F-theoretically a vev 
of an $SO(8)$ adjoint scalar field $\Phi$ living on D7-branes, say, 
\begin{eqnarray}
\langle \Phi \rangle=\left(
\begin{array}{cccc}
i\sigma_2 m_1&&&  \\
&i\sigma_2 m_2&& \\
&&i\sigma_2 m_3 & \\
&&&i\sigma_2 m_4 \\
\end{array}
\right), ~~~~~~~~~~
\sigma_2=\left(
\begin{array}{cc}
0&-i  \\
i&0\\
\end{array}
\right) ,
\end{eqnarray}
$\mu$'s (when $Q=0$) just stand for its diagonal 
Cartan elements 
w.r.t. $SO(4)\times SO(4)$ decomposition. 
In addition, $\mu$'s get related to 
momenta of 2D Liouville primary fields 
$V_\mu=e^{2\mu \varphi}$ ($\varphi$: Liouville field) 
with a conformal dimension 
$\Delta(\mu)=\mu(Q-\mu)$. A zero background charge $Q=b+ \dfrac{1}{b}=\dfrac{\epsilon_1 + \epsilon_2}{\sqrt{\epsilon_1 \epsilon_2}}=0$ required here 
corresponds to a 4D physical field theory limit  $\epsilon_1=-\epsilon_2$%
\footnote{$\epsilon_{1,2}$ are related to the size 
of a unit rectangle in Young tableaux appearing in Nekrasov's partition functions.}.

While these arguments do render a satisfactory 
explanation of the geometric origin of triality, we instead 
would like to explore another possibility using 
Gauss hypergeometric functions. 
What makes this accessible is again 
due to AGT conjecture 
which proposes an equivalence between a 
2D Liouville 
conformal block ${\cal B}$ defined on $C_{g,n}$ 
and the instanton part of Nekrasov's 
partition function $Z_{\text {inst}} [C_{g,n}]$ of 
a 4D $\N=2$ $A_1$-type SCFT. Under special 
circumstances, the four-point spherical 
${\cal B}[C_{0,4}]$ 
satisfies a hypergeometric differential equation (HDE). 
Therefore, based on the equality $Z_{\text {inst}} [C_{0,4}]
={\cal B}[C_{0,4}]$ with $q_{UV}$ regarded as the 
cross-ration of four punctures, one can interpret 
\eqref{ma1} and \eqref{ma} as 
interchanging solutions of a HDE 
fixed by some Riemann scheme because 
$Z_{\text {inst}}({\bf a}, \vec{m}, q_{UV}(\tau_0),
\epsilon_{1},\epsilon_{2})$ (${\bf a}$: 
Coulomb phase parameter) itself 
contains the exact solution (SW curve) to 
infra-red dynamics. 
To conclude, we find 
that triality generates six out of 
Kummer's twenty-four solutions. 
By grouping them properly into three pairs, each pair 
just spans the basis of solutions belonging to 
respectively $(0,1,\infty)$ 
known as regular singularities of a second-order HDE.

This letter is organized as 
follows. 
In section 2, we review necessary aspects 
about hypergeometric functions. 
Especially, the elliptic $lambda$ function relating 
$\tau_{UV}$ to $\tau_0$ 
will play 
a quite profound role in latter 
discussions. In section 3, we show how triality can be interpreted as 
the interplay among six of Kummer's twenty-four solutions 
via AGT conjecture. 
A summary is given in section 4.

\section{Preliminaries}

\subsection{Hypergeometric function }
Let us first recall some main features of Gauss hypergeometric functions and 
their relation to the modular curve 
$X_2={\mathbb H}/\Gamma(2)$ being isomorphic to 
${\mathbb C} \backslash \{ 0,1\}$. See for example \cite{MY} for 
details. Here, $X_N$ is in general a 
noncompact Riemann surface whilst ${\mathbb H}$ 
stands for the upper half-plane. $\Gamma(N)$ denotes the 
level $N$ principal congruence subgroup of $SL(2,\mathbb{Z})$:  
\begin{eqnarray*}
\Gamma(N) \ni \left(
\begin{array}{cc}
 a& b  \\
 c & d  
\end{array}
\right)\equiv 
\left(
\begin{array}{cc}
 1& 0  \\
 0 & 1  
\end{array}
\right) ~ ~ {\text {mod}}~ ~ N, ~ ~~ ~~ ~ad-bc=1.
\label{}
\end{eqnarray*}
Getting familiar with these stuffs serves as our cornerstone of clarifying 
the role of triality observed 
in the $2N_c=N_f=4$ SW theory 
by means of Gauss hypergeometric functions $y(z)$, solutions of 
a second-order linear ODE: 
\begin{eqnarray}
z(1-z)y(z)'' + \big( c-(a+b+1)z\big) y(z)' -aby(z)=0, ~ ~~ ~~ 
z\in \mathbb{C}.
\label{gh}
\end{eqnarray}
Meanwhile, there are three $regular$ singularities 
$(0,1,\infty)$ near each of which two linearly 
independent solutions to \eqref{gh} exist%
\footnote{At points other than $(0,1,\infty)$, 
\eqref{gh} can be simplified to $y(z)''=0$ by changes of variables.}. That is, at $z=0$
\begin{eqnarray*}
\begin{cases}
y_{01}= {}_2F_1 (a,b,c;z), \\
y_{02}= z^{1-c}{}_2F_1 (a-c+1, b-c+1,2-c;z);
\end{cases}
\end{eqnarray*}
at $z=1$
\begin{eqnarray*}
\begin{cases}
y_{11}= {}_2F_1 (a,b,a+b-c+1;1-z) , \\
y_{12}= (1-z)^{c-a-b}{}_2F_1 (c-a , c-b, c-a-b+1; 1-z);
\end{cases}
\end{eqnarray*}
at $z=\infty$
\begin{eqnarray*}
\begin{cases}
y_{\infty 1}= (-z)^{-a} {}_2F_1 (a,a-c+1,a-b+1;z^{-1}), \\
y_{\infty 2}= (-z)^{-b} {}_2F_1 (b, b-c+1, b-a+1;z^{-1}).
\end{cases}
\end{eqnarray*}
According to the local exponent of $z$ around each singularity, 
three pairs of solutions listed above can be summarized by 
Table \ref{t1} (Riemann scheme) in the context of Fuchsian linear differential equations. 
Also, due to Fuchs relation summing up all 
entries inside the last two rows of 
Table \ref{t1} gives zero. 
\begin{table}[h]
\caption{Riemann scheme}
  \label{t1}
  \begin{center}
    \begin{tabular}{|c|c|c|} \hline
$z=0$& $z=1$& $z=\infty$ \\ \hline
0&0&$a$\\ \hline
$1-c$& $c-a-b$ & $b$ \\ \hline
    \end{tabular}
  \end{center}
\end{table}
As a matter of fact, each pair of solutions 
can be transformed to one another 
through suitable two by two matrices (connection coefficients); for instance,   
\begin{eqnarray}
\label{cc}
&&(y_{01}, y_{02})= (y_{11}, y_{12})P_{01}, \nonumber \\
&&P_{01} =\left(
\begin{array}{cc}
 \dfrac{\Gamma(c)\Gamma(c-a-b)}{\Gamma(c-a)\Gamma(c-b)}& 
 \dfrac{\Gamma(2-c)\Gamma(c-a-b)}{\Gamma(1-a)\Gamma(1-b)}  \\
 \dfrac{\Gamma(c)\Gamma(a+b-c)}{\Gamma(a)\Gamma(b)}
 &\dfrac{\Gamma(2-c)\Gamma(a+b-c)}{\Gamma(a-c+1)\Gamma(b-c+1)}   
\end{array}
\right). 
\end{eqnarray}
Furthermore, 
$y(z)$ suffers monodromies like  
\begin{eqnarray*}
&&(y_{\ell 1}, y_{\ell 2})\to (y_{\ell 1}, y_{\ell 2})M ,
~~~~~~M \in \pi_1 ({\mathbb C} \backslash \{ 0,1\})
\end{eqnarray*}
when winding around each singularity. 
That there exists a group homomorphism between 
the fundamental group 
$\pi_1 ({\mathbb C} \backslash \{ 0,1\})$ and 
$GL(2, {\mathbb{C}})$ leads to 
$M(\gamma, z_0)\in GL(2, {\mathbb{C}})$ (modulo conjugation) 
for each path $\gamma$ given a reference point $z_0$. 
Due to $\gamma_1 \cdot \gamma_2=\gamma_\infty$ one is able to 
establish $M(\gamma_1, z_0) \cdot M(\gamma_2, z_0)=
M(\gamma_\infty, z_0)$ 
where 
$\gamma_1$ $(\gamma_2)$ is designated to 
surround the singularity $z=0$ ($z=1$) 
counterclockwise.

\subsection{ Schwarz map}

Next, we proceed to consider 
the ratio 
${\cal D}_\ell=\dfrac{y_{\ell 1}}{y_{\ell 2}}$ 
defining the famous triangle Schwarz map, a special 
case of conformal Schwarz-Christoffel maps which 
bring the upper half-plane ${\mathbb H}$ to certain 
$n$-vertex polygon. In fact, the setup under consideration can be cast into 
the so-called uniformization problem for the simplest case--
a three-punctured sphere ${\mathbb C} \backslash \{ 0,1\}$. 
One can arrange \eqref{gh} into a $Q$-form 
\begin{eqnarray*}
\partial_z^2 y + \dfrac{1}{2}\{ \rho ,z \}y=0, ~~~~~~
\{ \rho ,z \}:~ \text {Schwarzian derivative of}~ \rho, ~~~~~~\rho=\dfrac{y_\varrho}{y_\varsigma}.
\end{eqnarray*}
There, the multi-valued $\rho$ (ratio of two independent solutions ) induces 
a map ${\mathbb C} \backslash \{ 0,1\} \to$ (unit disc)
/${ \cal G}$ with branching points $(0,1,\infty)$. 
${\cal G} \subset SU(1,1)$ 
denotes the monodromy group of 
$\rho$ as will soon be seen.

While one takes $\rho={\cal D}_{\ell}$, it naively 
maps ${\mathbb H}$ 
to a triangle on a Riemann sphere ${\mathbb P}^1$ bounded by circular arcs. Connection coefficients $P$'s can thus be thought of as applying 
M{\"o}bius transformations 
(automorphism group of ${\mathbb P}^1$) to the triangle. 
Meanwhile, $(0,1,\infty)$ on ${\mathbb H}$ 
are brought to three vertices whose angles are 
$\pi \nu_\ell $ respectively:   
\begin{eqnarray}
\label{nu}
\nu_0= 1-c=\dfrac{1}{p}, ~~~~~~
\nu_{1}= c-a-b=\dfrac{1}{q}, ~~~~~~
\nu_{\infty}= b-a=\dfrac{1}{r} .
\end{eqnarray}
$(p,q,r)$ are natural numbers greater than one. 
The relation between \eqref{nu} and Table \ref{t1} can be made clear if one looks into the local behavior of ${\cal D}_\ell$ 
near each responsible singularity:
\begin{eqnarray*}
\label{}
{\cal D}_0 \simeq z^{\nu_0} \big( 1+{\cal O}(z)\big), ~~~~
{\cal D}_1 \simeq (1-z)^{\nu_1} \big( 1+{\cal O}(1-z)\big), ~~~~
{\cal D}_{\infty} \simeq z^{-\nu_{\infty}} 
\big( 1+{\cal O}(z^{-1})\big) .
\end{eqnarray*}
Of course, extending ${\cal D}_\ell({\mathbb H})$ to 
${\cal D}_\ell({\mathbb C})$ 
is totally possible and one encounters  
\begin{eqnarray}
\label{mono}
{\cal D}_{\ell} \to \dfrac{a {\cal D}_{\ell} + b}{c 
{\cal D}_{\ell} + d}, 
~~~~~~~~~~~\left(
\begin{array}{cc}
 a& b  \\
 c & d  
\end{array}
\right)\in GL(2,\mathbb{C})
\end{eqnarray}
just because of monodromies when winding around each responsible singularity. The resulting image 
${\cal D}_{\ell} ({\mathbb C})$ becomes two sets of triangles%
\footnote{For instance, one can paint each triangle certain 
color according to which one of two half-planes they come from.} which 
tile the entire ${\mathbb P}^1$ if \eqref{nu} is assumed.  Certainly, after an automorphism (pattern-preserving) group $\Gamma(p,q,r)$ is 
divided, one is able to claim that 
\begin{eqnarray}
{\cal D}_\ell: ~{\mathbb C} \backslash \{ 0,1\} \to  \big( {\mathbb P}^1-{  
{\text {triangle~vertices}}} \big)/\Gamma(p,q,r).
\label{MO}
\end{eqnarray}

As will be justified below, 
in view of \eqref{mono} we cannot help but 
regard ${\cal D}_\ell$ as the complex moduli of some 
elliptic curve $E_{{\cal D}_\ell}$ with 
${\cal D}_\ell\in {\mathbb H}/g$ (modular curve). 
This way of thinking is also inspired by 
the definition of ${\cal D}_\ell$ being a ratio of two 
hypergeometric functions both of which solve 
Fuchsian equations and are 
identified with period integrals over 
an algebraic curve. That generators of $g$ must be those of 
$\pi_1 ( {\mathbb C} \backslash \{ 0,1\},z_0 )$ w.r.t. 
${\cal D}_\ell$ confirms that there exists 
a group homomorphism between 
$\pi_1 ( {\mathbb C}\backslash \{ 0,1\},z_0 )$, 
$\Gamma(p,q,r)$ and $g$. Indeed, we will 
find that the above picture is realized 
when $(p,q,r)=(\infty, \infty, \infty)$ 
and $g=\Gamma(2)$.

\subsection{Elliptic curve and $\lambda$-function}
Let us proceed to clarify the appearance of an elliptic curve $E_{\cal D}$ mentioned above.  
An integral representation of 
${}_2F_1 (a,b,c;z)$ 
for $z\ne (0, 1, \infty)$ is given by%
\footnote{Usual normalizations 
like Beta factors are omitted. Note also that 
with 
$(z_1, z_2, z_3,z_4)=(0,1,z,\infty)$ a simplification occurs, i.e.  
\begin{eqnarray*}
\label{}
\int_{\gamma}  \prod^4_{i=1} (u-z_i)^{-\mu_i} du~~~\to ~~~
  \int_{\gamma}  \prod^3_{i=1} (u-z_i)^{-\mu_i} du
\end{eqnarray*}
due to the term involving $z_4=\infty$ dropped. }
\begin{eqnarray}
\label{ig}
\int_{\gamma} u ^{-\mu_0} (u-1)^{-\mu_1} (u-z)^{-\mu_z} du 
= \int_{\gamma} \eta(z), ~~~~~~~~\mu_0+ \mu_1+ \mu_z+ \mu_{\infty}=2
\end{eqnarray}
where all $\mu$'s are simple linear combinations of 
$(a,b,c)$ and assumed to be rational. 
$\eta(z) \equiv \dfrac{du}{x}$ is defined w.r.t. an 
algebraic curve 
\begin{eqnarray}
\label{ac}
X: ~x^\kappa = u^{\kappa\mu_0} (u-1)^{\kappa\mu_1} (u-z)^{\kappa\mu_z}
\end{eqnarray}
with 
$\kappa$ being the least common denominator of $\mu$'s. 
$\gamma$ known as Pochhammer's contour now becomes 
some homology cycle of $X$, i.e. 
$\gamma\in H_1 (X, {\mathbb Z})$. Inequivalent $\gamma$'s will 
lead to independent hypergeometric functions. 
$X$ turns out to be an elliptic curve 
of the standard Legendre form: 
$x^2=4u(u-1)(u-z)$ 
when $\kappa=2$ and $\mu_0= \mu_1= \mu_z= \mu_{\infty}=\frac{1}{2}$. This soon implies 
$(a,b,c)=(\frac{1}{2},\frac{1}{2},1)$%
\footnote{Equivalently, 
$\nu_0= \nu_1= \nu_{\infty}=0$ or 
$p=\lambda=r=\infty$.} from the parameterization:
\begin{eqnarray*}
\begin{cases}
\label{}
\mu_0=\frac{1}{2} (1-\nu_0+\nu_1 -\nu_{\infty}), ~~~~~~~~
\mu_{1}=\frac{1}{2} (1+\nu_0-\nu_1 -\nu_{\infty}),\\
\mu_{z}=\frac{1}{2} (1-\nu_0-\nu_1 +\nu_{\infty}), ~~~~~~~~
\mu_{\infty}=\frac{1}{2} (1+\nu_0+\nu_1 +\nu_{\infty}). \\
\end{cases}
\end{eqnarray*}
Now, consider the ratio 
\begin{eqnarray}
\label{ddd}
\tilde{\cal D}(z)=\dfrac{{}_2F_1 (\frac{1}{2},\frac{1}{2},1;z)}{{}_2F_1 (\frac{1}{2},\frac{1}{2},1;1-z)}
=\dfrac{K(\sqrt{z})}{K'(\sqrt{z})} , ~~~~~~~
K'(\sqrt{z})=K(\sqrt{1-z})
\end{eqnarray}  
where $K$ denotes the complete elliptic integral 
of the first kind. Conventionally, $\tilde{\cal D}(z)/2$ 
is called the $aspect$ ratio of a rectangle 
yielded by performing 
a Schwarz-Christoffel map 
over ${\mathbb H}$. 
Through defining 
\begin{eqnarray*}
\label{}
\tau\equiv i\tilde{\cal D}(z)
=\dfrac{\int_{\gamma_1} \eta(z)}{\int_{\gamma_2} \eta(z)}, 
~~~~~~\gamma \in H_1 \big(X(\mu_i), {\mathbb Z}\big), 
~~~~~~\forall\mu_i={1}/{2},
\end{eqnarray*} 
the famous isomorphism: 
\begin{eqnarray}
\label{iso}
 i\tilde{\cal D}: ~\mathbb{C} \backslash \{0,1\}\to  {\mathbb H}/\Gamma(2)
\end{eqnarray}
is induced. In particular, 
the appearance of $\Gamma(2)$ is due to the choice of 
$(a,b,c)$ as explained around \eqref{mmm}. 
${\tau}$ becomes exactly 
the complex moduli of a torus 
${\mathbb C}/\Lambda_{\tau}$ ($\Lambda_{\tau} \equiv 
{\mathbb Z}{\tau}+ {\mathbb Z}$) which is 
isomorphic to $X$ in \eqref{ac} with 
$\forall\mu_i={1}/{2}$. 
In addition, the inverse of 
$\tilde{\cal D}(z)$ is known as 
the elliptic $lambda$ function: 
\begin{eqnarray}
\label{la}
\lambda\equiv z=\frac{\theta_2^4 (q)}{\theta_3^4 (q)}
\end{eqnarray} 
where $\theta_i(q)$'s are theta constants whilst 
$q=e^{i\pi \tau}=e^{-\pi \tilde{\cal D}}$ is called the $nome$. 
From now on, we will not especially 
distinguish between 
$\lambda$ and $z$ which eventually represent 
the cross-ratio of four points on ${\mathbb P}^1$%
\footnote{The cross-ratio of four points on 
${\mathbb P}^1$ is given by \begin{eqnarray}
\lambda=(x_2, x_1; x_3,x_4)=\frac{(x_2-x_3)(x_1-x_4)}{(x_1-x_3)(x_2-x_4)}.\nonumber
\end{eqnarray} }. 
By definition $\lambda$ should be invariant under $\Gamma(2)$ 
or, equivalently, subject to 
\begin{eqnarray}
\lambda(\tau+2)=\lambda(\tau)
,~~~~~~~~~~~\lambda\Big( \frac{\tau}{1-2\tau} 
\Big)=\lambda(\tau).\nonumber 
\end{eqnarray}

Furthermore, since \eqref{iso} is a bijective map 
what is inferred is that 
six distinct $\lambda$'s 
define the same elliptic curve 
because of same 
Klein's absolute $j$-invariants they will provide. Namely, 
a six-to-one relation does follow owing to   
\begin{eqnarray}
j=\frac{4}{27}\frac{\big(1-\lambda+\lambda^2 \big)^3}{\lambda^2 
\big(1-\lambda \big)^2}=\frac{g_2^3}{g_2^3 - 27g_3^2}
\label{jj}
\end{eqnarray}
where $g_2$ and $g_3$ are 
modular invariants of an 
elliptic curve expressed in Weierstrass form: 
\begin{eqnarray*}
y^2=4x^3 -g_2 x -g_3 
\end{eqnarray*}
whose three distinct roots are $(e_1,e_2,e_3):=\Big(  
\wp(\frac{1}{2}), \wp(\frac{\tau}{2}),\wp(\frac{\tau}{2}+\frac{1}{2}) 
\Big)$. Notice that $\wp$ is Weierstrass's doubly-periodic 
function. Consequently, 
\begin{eqnarray*}
\lambda=\frac{\wp(\frac{\tau}{2}+\frac{1}{2})-\wp(\frac{1}{2})}{
\wp(\frac{\tau}{2})-\wp(\frac{\tau}{2}+\frac{1}{2})}.
\end{eqnarray*}

We are led to the following conclusion. 
Upon defining Hom$\big( \pi_1( 
\mathbb{C} \backslash \{0,1\}, z_0), SL(2, \mathbb{Z}) \big)$, because generators of 
the monodromy group w.r.t. 
${}_2 F_1 (a,b,c;z)$ are determined 
by $(a,b,c)$ as 
\begin{eqnarray*}
&&M_1=\left(
\begin{array}{cc}
 1& 0  \\
 -1+e^{-2\pi i b} & e^{-2\pi ic}  
\end{array}
\right) ~~~{\text {and}}~~~\nonumber\\
&&M_2=\left(
\begin{array}{cc}
1& 1-e^{-2\pi ia}  \\
 0 & e^{-2\pi i(a+b-c)} 
\end{array}
\right)~~~{\text {with}}~~~
(a, b, c-a ,c-b) \notin {\mathbb Z},
\end{eqnarray*}
when $(a,b,c)=(\frac{1}{2},\frac{1}{2},1)$ they 
are just those of $\Gamma(2)$, i.e.  
\begin{eqnarray}
\label{mmm}
M_1 =\left(
\begin{array}{cc}
 1& 0  \\
 -2& 1  
\end{array}
\right) ~~~{\text {and}}~~~
M_2=
\left(
\begin{array}{cc}
 1& 2  \\
 0&1  
\end{array}
\right)
\end{eqnarray}
as used in \eqref{iso}. 
As a remark, the relation 
\eqref{ddd} is completely not new 
since it has long been known as the 
infra-red gauge coupling 
$\tau_{IR}$ in the pure $SU(2)$ 
SW theory if one equates 
${(2-z)}/{z}$ with its Coulomb phase parameter there.

All in all, we have just wandered quite a lot from the conventional 
interpretation of 
\eqref{iso}, i.e. one can 
always express an elliptic curve 
in terms of a two-sheeted cover of 
a sphere with branching points $(0,1,\lambda,\infty)$ 
such that the equivalence between their moduli spaces 
naturally introduces the underlying isomorphism 
\eqref{iso} or its inverse--$\lambda$-function \eqref{la}.

\section{Triality in $SU(2)$ $N_f=4$ Seiberg-Witten theory}
A standard $2N_c=N_f=4$ SW curve 
is of a rather 
sophisticated form 
parameterized by 
four bare flavor masses and $\tau_0$ \cite{Seiberg:1994aj}:  
\begin{eqnarray}
y^2=4\Big[W_1W_2W_3+A\big(W_1T_1(e_2-e_3)+W_2T_2(e_3-e_1) +W_3T_3(e_1-e_2)\big)-A^2 N \Big] 
\label{swc}
\end{eqnarray}
where ($u$: Coulomb phase parameter)
\begin{eqnarray*}
{W_i}=x-{e_i}u-{{e_i}}^{2}R,~~~~~
A= \left( {e_1}-{e_2} \right)  \left( {e_2}-{e_3} \right) 
 \left( {e_3}-{e_1} \right)
\end{eqnarray*}
and
\begin{eqnarray*}
&&R=\dfrac{1}{2}\sum_{i=1}^4m_i^2, ~~~~~
N=\dfrac{3}{16}\sum_{i>j>k}m_i^2m_j^2m_k^2-\dfrac{1}{96}
\sum_{i\neq j}m_i^2m_j^4+\dfrac{1}{96}\sum_{i=1}^4m_i^6,
\nonumber\\
&&T_1=\dfrac{1}{12}\sum_{i>j}m_i^2m_j^2-\dfrac{1}{24}\sum_{i=1}^4m_i^4,  ~~~~~T_2=-\dfrac{1}{2}\prod_{i=1}^4 m_i-\dfrac{1}{24}\sum_{i>j}m_i^2m_j^2+\dfrac{1}{48}\sum_{i=1}^4m_i^4,\nonumber\\
&&T_3=\dfrac{1}{2}\prod_{i=1}^4m_i-\dfrac{1}{24}\sum_{i>j}
m_i^2m_j^2+\dfrac{1}{48}\sum_{i=1}^4m_i^4.
\nonumber
\end{eqnarray*} 
$e_i$'s are functions of $\tau_0$: 
\begin{eqnarray*}
e_1=\dfrac{1}{12}({\theta}_3^4+\theta_4^4),~~~~~ e_2=\dfrac{1}{12}(\theta_2^4-\theta_4^4), ~~~~~
 e_3=\dfrac{1}{12}(-\theta_2^4-\theta_3^4).\nonumber
\end{eqnarray*}
Here, 
$\tau_0$ must be regarded as the asymptotic value of 
$\tau_{IR}= \tau_0 +\dfrac{1}{2\pi i}\Big( 
\sum_{i=1}^4 \log(u-m_i^2) 
-4\log u \Big)+\cdots$ expanded at large 
$u$. Its reduction to asymptotically-free 
counterparts ($N_f\le 3$)
is easily achieved via tuning $\tau_0$ and $m_i$ 
in order to yield a suitable dynamical scale $\Lambda_{N_f}$.

Seiberg and Witten found that \eqref{swc} 
is invariant under elements of 
\begin{eqnarray}
SL(2,\mathbb{Z})/\Gamma(2)=\mathbf{S}_3=\{ I, ~S, ~T, 
~ST, ~TS, ~STS \}
\label{TT}
\end{eqnarray}
if and only if \eqref{ma1} and \eqref{ma} 
are taken into account simultaneously. This phenomenon 
is often referred to as triality whose origin may 
be owing to 
the outer automorphism group ${\mathbf S}_3$ of 
$Spin(8)$, the quantum 
flavor symmetry in the superconformal case ($m_i=0$). 
Because we want to interpret triality 
as interchanging Kummer's solutions, 
our strategy is to think of 
$\Gamma(1)/\Gamma(2)=\mathbf{S}_3$ here as 
$\mathbf{S}_4/({\mathbb Z}_2 \times {\mathbb Z}_2 )$ on 
a four-punctured ${\mathbb P}^1$ via $\lambda$-function introduced in \eqref{la}%
\footnote{In AGT's Appendix (B.29) 
$\lambda$-function has already shown up. }.

In other words, under $\mathbf{S}_3$ $\tau_0$ enlarges 
its ``fundamental" domain%
\footnote{As \eqref{swc} reduces to merely an usual 
Weierstrass elliptic curve characterized by $\tau_0$ 
when all $m_i=0$, so basically $\tau_0 \in {\mathbb H}/SL(2,{\mathbb Z})$.} instead to 
${\mathbb H}/\Gamma(2)$ 
and by \eqref{la} we will now map it 
bijectively 
to $\lambda$-space defined on 
$C_{0,4}$ where 
six distinct cross-ratios are 
caused by applying $\mathbf{S}_4/({\mathbb Z}_2 \times {\mathbb Z}_2 )$. That is, the action of ${\bf S}_3$ 
is translated to interchanging four marked punctures. 
The next step is to know how 
\eqref{ma1} and \eqref{ma} can be 
incorporated 
into the four-point spherical conformal block 
${\cal B}[C_{0,4}]$ 
with 
\begin{eqnarray}
{\cal B}
(\alpha, \vec{\mu}, Q|\lambda)=
Z^{2N_c =N_f =4}_{inst} ({\bf a}, \vec{m},  \epsilon_{1,2}|
q_{UV}),  
~~~~~~
\alpha: ~\text {internal ~momentum}
\label{ttt}
\end{eqnarray}
where in addition to $\lambda\equiv q_{UV}(\tau_0)$ the 
dictionary between parameters on two sides 
has been spelt out by AGT. 
All in all, making use of properties of 
Gauss hypergeometric 
functions we will arrive at a new unifying 
understanding of this mysterious 
part of $S$-duality--triality for $2N_c=N_f=4$.

\subsection{Gaiotto's picture and AGT conjecture}
Gaiotto's idea arises from rearranging old SW curves and 
leads to another way of engineering a huge class of $\N=2$ $SU(r)$ SCFTs by wrapping $r$ M5-branes on $C_{g,n}$, i.e. compactifying 
6D $A_{r\text{-}1}$-type $(2,0)$ theories 
on $C_{g,n}$ accompanied by a partial twisting. 
There are various ways of decomposing 
$C_{g,n}$ into trinions and tubes so 
$S$-duality (mapping class) group gets identified with such 
physically equivalent surgeries. In addition, 
weak-coupling limits are attained intuitively 
by elongating extremely tubes joining two 
punctures. This kind of 
$degenerate$ limits correspond to $cusps$ 
in the moduli space of $C_{g,n}$. 
Total $3g$-3+$n$ tubes contained in 
$C_{g,n}$ correspond to 
the number of gauge groups of a weakly-coupled 
quiver SCFT equipped with a Lagrangian description.

Let us elaborate arguments about aforementioned 
$\lambda$-space on $C_{0,4}$. Upon viewing $\lambda$ as the coordinate on ${\mathbb P}^1 \backslash 
(0,1,\infty)$ (up to a M{\"o}bius transformation), 
six distinct values generated from it by 
$\Gamma(1)/\Gamma(2)= \mathbf{S}_3$ are 
referred to as six different cross-ratios: 
\begin{eqnarray}
\begin{array}{cccccccc}
\mathrm{element} ~\mathrm{in}~ \mathbf{S}_3  &&I &T &S &ST &TS &STS   \\ 
\mathrm{cross}\text{-}\mathrm{ratio}&&\lambda&\dfrac{\lambda}{\lambda-1}&1-\lambda&\dfrac{1}{1-\lambda}&\dfrac{\lambda-1}{\lambda}&\dfrac{1}{\lambda}\\ 
\end{array}
\label{6q}
\end{eqnarray}
\eqref{6q} can be 
derived directly based on 
either footnote 10 or \eqref{la} with modular properties 
of theta constants listed below: 
\begin{eqnarray*}
&&\theta_{2}(q)\equiv \vartheta_{10} (0,\tau) = \frac{1}{\sqrt{-i\tau}}
\vartheta_{01} (0,-\frac{1}{\tau}), ~~~~~~
\theta_{3}(q) \equiv\vartheta_{00} (0,\tau) = \frac{1}{\sqrt{-i\tau}}
 \vartheta_{00} (0,-\frac{1}{\tau}),\nonumber\\
&&\theta_4 (q)\equiv \vartheta_{01} (0,\tau) = 
\vartheta_{00} (0,{\tau+1}), ~~~~~~~~~~~~
\vartheta_{10} (0,\tau) = e^{- \frac{i\pi}{4}} 
 \vartheta_{10} (0,{\tau +1}).
\end{eqnarray*}
Replacing 
${\mathbb H}/\Gamma(2)$ by 
$\Big( {\mathbb H}/ SL(2, \mathbb{Z})\Big) \times {\mathbf S}_3$ in 
\eqref{iso} and recalling 
that another famous 
isomorphism 
\begin{eqnarray*}
\label{}
j: ~{\mathbb H}/ SL({2, \mathbb{Z}})
 \to {\mathbb C} \backslash \{0,1\}  
\end{eqnarray*}
is induced by Klein's $j$-invariant, one agrees 
that 
the identity \eqref{jj} between $j$ and $\lambda$ describes a six-to-one relation.

Next, we focus on ${\cal B}[C_{0,4}]$ in \eqref{ttt}. 
In general, because it satisfies 
Zamolodchikov's recursion relation \cite{Zamolodchikov:1995aa} 
an expansion over $\lambda$ 
to any desired order is possible. However, 
a closed-form expression of it is still missing. 
Nevertheless, if 
one of four inserted primary fields 
becomes degenerate, say, $V_{\mu_3}=\Phi_{2,1}$ it is well-known 
\cite{BPZ, Fateev:2009me} that 
by means of the null-state condition: 
\begin{eqnarray*}
\Big( L_{-2} -\frac{3}{2\big( 
2\Delta(h_{2,1}) +1 \big)}L^2_{-1} \Big) 
\Phi_{2,1}=0 , 
~~~~~~  h_{r,s}=\frac{1-r}{2}b + 
\frac{1-s}{2b},~~~~~~ \mu_3\equiv h_{2,1}=-\dfrac{b}{2},
\end{eqnarray*}
one is led to 
\begin{eqnarray}
\label{cbhy}
{\cal B}[C_{0,4}]\equiv \langle V_{\mu_1}(0) V_{\mu_2}
(1) V_{\mu_3}(\lambda) V_{\mu_4}(\infty)\rangle=
\lambda^{b\mu_1} (1-\lambda)^{b\mu_2} {}_2F_1(a,b,c;\lambda)
\end{eqnarray}
where%
\footnote{We adhere to conventions used in 
\cite{Schiappa:2009cc}. Also, we wish 
``$b$" used 
in both ${}_2F_1(a,b,c; z)$ and Liouville theory side 
will cause no confusion.} 
\begin{eqnarray}
\label{abcn}
\begin{cases}
a=-N,\\
b=\dfrac{1}{\beta}(-\dfrac{2\mu_1}{\epsilon_1}
-\dfrac{2\mu_2}{\epsilon_1}+2) +N-1,\\
c=\dfrac{1}{\beta}(-\dfrac{2\mu_1}{\epsilon_1}+1),\\
N=-\epsilon_1(\mu_1+\mu_2+\mu_3-\mu_4),~~~~~
\beta=-\dfrac{\epsilon_2}{\epsilon_1}, ~~~~~\epsilon_1=b, ~~~~~
\epsilon_2=\dfrac{1}{b}.
\end{cases}
\end{eqnarray} 
The internal momentum $\alpha$ is set 
to be 
\begin{eqnarray*}
\alpha=\frac{Q}{2}+{{\bf a}}=\mu_{4}+\frac{b}{2}, ~~~~~~~~~~
Q=b + \frac{1}{b}. 
\end{eqnarray*}
Adopting a 4D physical field theory limit 
$\epsilon_{1}+\epsilon_2=0$ ($\beta=1$) may give rise to 
a further simplification%
\footnote{$\epsilon_{1}=-\epsilon_2$ serves as the 
$genus$-expansion parameter inside 
$Z_{{\text{inst}}}=\exp({{\cal F}})$ since 
${\cal F}=-\frac{1}{\epsilon_{1}\epsilon_{2}}{\cal F}_0+\cdots$ is referred to as 
the A-model topological string free energy 
w.r.t. a responsible 
Calabi-Yau three-fold in Type IIA theory.}. 
Note that $N$ is designated to characterize 
the size of a $hermitian$ matrix appearing in 
the recent Dijkgraaf-Vafa proposal \cite{Dijkgraaf:2009pc}. 
There, an ($A_1$-type) $n$-point spherical 
${\cal B}$ was rewritten in terms of a Penner-type 
matrix integral 
(or Selberg-Kaneko integral \cite{ka}). 
As a remark, 
from \eqref{abcn} ${}_2F_1(a,b,c;\lambda)$ also 
stands for 
a Jacobi polynomial defined by 
\begin{eqnarray*}
&&G_N(\xi, \zeta;\lambda)={}_2F_1(-N,\xi+N, \zeta;\lambda)
=1+\sum_{r=1}^N (-)^r {}_N {\text C}_r
\dfrac{\Gamma(\xi+N+r)\Gamma(\zeta)}{\Gamma(\xi+N)\Gamma(\zeta+r)}\lambda^r, \\
&&\text{for} ~
\zeta\ne 0, -1, -2, \cdots,-N+1.  
\end{eqnarray*}
This sounds quite consistent with the fact pointed out by 
Schiappa and Wyllard \cite{Schiappa:2009cc} 
that a three-point DV matrix model $Z^{DV}_{3pt}$ 
for ${\cal T}_{0,3}(A_1)$ is 
exactly solved by its orthogonal polynomial--Jacobi 
polynomial; namely, 
$\langle \det(\lambda-M)\rangle_{Z^{DV}_{3pt}}$ is equal to 
\eqref{cbhy} without the factor 
$\lambda^{b\mu_1} (1-\lambda)^{b\mu_2}$ as shown in \cite{Schiappa:2009cc}.

\subsubsection{Relation to ${\cal N}=2^\ast$ $A_1$ system}
Inspired by the appearance of a Jacobi 
polynomial said above, we strongly 
expect that its reduction to $A_1$-Jack 
(or Gegenbauer) polynomials by further 
constraining three $\mu$'s can be given 
a physical interpretation%
\footnote{I thank Hirotaka Irie, Yutaka Matsuo and 
Akitsugu Miwa with whom I have communicated about these stuffs.}. 
Namely, having in mind that an $A_1$-Jack polynomial 
gets closely related to a specialized hypergeometric function 
\begin{eqnarray}
{}_2F_1(-A,A+2B,B+\frac{1}{2};x)  
\label{jjjj}
\end{eqnarray}
and is the eigenstate 
of $A_1$-type Calogero-Sutherland model, 
a limiting case of 
$A_1$-type Calogero-Moser model as 
$p=\exp(2\pi i \tau)$ of 
Weierstrass's $\wp$-function 
goes to zero (or $\tau \to i\infty$), we cannot help 
suspecting that the constraint imposed on 
three $\mu$'s leading to \eqref{jjjj} should result from 
a one-punctured $pinched$ torus. In other words, one may think of \eqref{jjjj} as a two-point conformal block ${\cal B}[C_{1,2}]$ with 
one insertion being $\Phi_{1,2}$ defined on a pinched torus.  Notice that redefining 
$x\sim \exp(i\lambda)\in{\mathbb C}^\ast$ 
makes the periodicity $\lambda\sim \lambda+2\pi$ explicit.

To carry out the check, one also needs to know the degenerating process: 
${\cal T}_{1,1}\to {\cal T}_{0,3}$. 
Given the fact that 
in the physical ${\cal T}_{0,3}$ theory four free hypermultiplets 
have their masses 
$\mu_1 \pm \mu_2 \pm \mu_4$ yielded 
from assigned momenta of 
three inserted Liouville primary fields \cite{Alday:2009aq}, 
plausibly $\mu$'s will now not be independent 
because in the former there are 
only two independent variables 
$({\mathfrak a}, {\mathfrak m})$, 
i.e. ${\cal N}=2^\ast$ $SU(2)$ Coulomb branch 
parameter and adjoint hypermultiplet mass. 
Further, from a toric diagram associated with 
a Calabi-Yau three-fold 
engineering $\N=2^{\ast}$ $SU(2)$ theory, 
one is able to read off masses of four free 
hypermultiplets in terms of 
$({\mathfrak a}, {\mathfrak m})$. Then, the constraint 
for $\mu$'s gained from comparing 
\eqref{cbhy} with \eqref{jjjj} can be directly contrasted 
with what is derived above via an $\N=2^\ast$ toric diagram.

An even interesting 
direction is to consider connections between various 
orthogonal polynomials by means of AGT picture. 
Serving as eigenstates of 
distinct Schr{\"o}dinger equations 
(or two-body integrable systems), 
they are nonetheless transformed to one another 
by performing some limit which may acquire 
suitable geometric meaning in terms of 
Riemann surfaces. 
We wish to report these topics in an upcoming paper.

\subsection{Triality and Kummer's 24 solutions} 
Mathematically speaking, multiplying prefactors like $\lambda^{A}(1-\lambda)^{B}$ just brings ${}_2F_1(a,b,c;\lambda)$ to 
another Riemann scheme containing solutions like 
${}_2F_1(a',b',c';\lambda)$ and so on. 
To say which scheme is more preferable seems 
not so essential. 
We decide to exclude 
these prefactors below also 
because in \cite{Fujita:2009gf} this choice of 
${\cal B}$ 
did reproduce the gravitationally-corrected 
asymptotically-free SW prepotential 
${\cal F}_0$. 

By applying AGT dictionary \eqref{mu0} together with 
\eqref{ma1}, \eqref{ma} and 
\eqref{abcn} specialized at $\beta=1$, it is seen that 
other five of Kummer's twenty-four solutions can be 
generated from 
${}_2F_1(a,b,c;\lambda)$ by 
elements $\{ S,T,TS,STS,ST \}$: 
\\
\\
$\underline{\mathbf{(1)~ S:} ~~  
\mathbf{\mu_1 \leftrightarrow \mu_2}}$ 
\begin{eqnarray*}
\begin{cases}
a\to a\\
b\to b\\
c\to a+b-c+1\\
\lambda \to 1-\lambda
\end{cases}
\end{eqnarray*}
$\underline{\mathbf{(2)~ T:} ~~ 
\mathbf{ \mu_4 \leftrightarrow \mu_2}}$ 
\begin{eqnarray*}
\begin{cases}
a\to -a+c\\
b\to b\\
c\to c\\
\lambda\to \dfrac{\lambda}{\lambda-1}
\end{cases}
\end{eqnarray*}
$\underline{\mathbf{(3)~ TS:} ~~ 
\mathbf{(\mu_1, \mu_2, \mu_4) \rightarrow (\mu_2, \mu_4, \mu_1)}}$ 
\begin{eqnarray*}
\begin{cases}
a\to b-c+1\\
b\to b\\
c\to a+b-c+1\\
\lambda\to \dfrac{\lambda-1}{\lambda}
\end{cases}
\end{eqnarray*}
$\underline{\mathbf{(4)~ STS:} ~~
\mathbf{ \mu_4 \leftrightarrow \mu_1}}$
\begin{eqnarray*}
\begin{cases}
a\to b+c-1\\
b\to b\\
c\to -a+b+1\\
\lambda\to \dfrac{1}{\lambda}
\end{cases}
\end{eqnarray*}
$\underline{\mathbf{(5)~ ST:} ~~
\mathbf{
 (\mu_1, \mu_2, \mu_4) \rightarrow (\mu_4, \mu_1, \mu_2)}}$
\begin{eqnarray*}
\begin{cases}
a\to -a+c\\
b\to b\\
c\to -a+b+1\\
 \lambda\to \dfrac{1}{1-\lambda}
\end{cases}
\end{eqnarray*}
Finally, all of them are collected below:  
\begin{eqnarray*}
\begin{cases}
\text{(I)}~{}_2F_1(a, b,c; \lambda)\\
\text{(II)}~{}_2F_1(a, b,a+b-c+1; 1-\lambda)\\
\text{(III)}~(1-\lambda)^{-b} {}_2F_1(c-a, b,c; \dfrac{\lambda}{\lambda-1})\\
 \text{(IV)}~\lambda^{-b} {}_2F_1(b-c+1, b,a+b-c+1; \dfrac{\lambda-1}{\lambda})\\
 \text{(V)}~ \lambda^{-b} {}_2F_1(b-c+1, b,b-a+1; \dfrac{1}{\lambda})\\
 \text{(VI)}~ (1-\lambda)^{-b} {}_2F_1(-a+c, b,b-a+1; \dfrac{1}{1-\lambda})
\end{cases}
\end{eqnarray*}
Although strictly speaking (III)--(V) have been 
dressed by either $\lambda^{b}$ or $(1-\lambda)^{b}$, 
it can merely be 
added by hand in order to preserve the given 
Riemann scheme. As a matter of fact, 
according to \cite{iwa} we see that 
\begin{eqnarray*}
\begin{cases}
\text{(I)~(III) }
\end{cases}
\lambda=0 ~{\text{basis,}}~~~~~~
\begin{cases}
\text{(II)~(IV) }
\end{cases}
\lambda=1 ~{\text{basis,}}~~~~~~
\begin{cases}
\text{(V)~(VI) }
\end{cases}
\lambda=\infty ~{\text{basis,}}~~
\label{51}
\end{eqnarray*}
where by ``basis" we mean spanning a basis of 
solutions around there. 
Therefore, triality plays the role of 
interchanging solutions around three regular singularities $(0,1,\infty)$. This 
manipulation thus manifests 
how triality is just understood in terms of 
another mathematical object--hypergeometric function.

\section{Discussion}

\subsection*{(I) Crossing symmetry and triality} 

Correlators $\Omega$ of 
some rational CFT including Liouville field theory (LFT) 
defined on a four-punctured Riemann sphere 
are endowed with the $crossing$ $symmetry$. 
Usually, this fact is expressed as follows: 
\begin{eqnarray} 
\Omega^{(4)} 
(
\begin{array}{cc}
\mu_2 & \mu_3   \\
\mu_4 & \mu_1  
\end{array}
\Big|\lambda)
=\Omega^{(4)} 
(
\begin{array}{cc}
\mu_1 & \mu_3   \\
\mu_4 & \mu_2 
\end{array}
\Big|1-\lambda)
\end{eqnarray}
where $\Omega^{(4)}$ is analytic over 
${\mathbb P}^1_{\ast}: {\mathbb P}^1 \backslash \{ 0,1,\infty\}$. 
When it comes to conformal blocks ${\cal B}$ 
(main component constituting the above $\Omega^{(4)}$) though, 
the crossing symmetry is respected to a $covariant$ extent: 
\begin{eqnarray} 
{\cal B}_{\mu_{31}}^{s} 
(
\begin{array}{cc}
\mu_2 & \mu_3   \\
\mu_4 & \mu_1  
\end{array}
\Big|\lambda)
=\int_{\frac{Q}{2}+i{\mathbb R}^+} d\mu_{23}~
F_{\mu_{31}\mu_{23}}^{L} 
\left(
\begin{array}{cc}
\mu_2 & \mu_3   \\
\mu_4 & \mu_1 
\end{array}
\right)
{\cal B}_{\mu_{23}}^{t} 
(
\begin{array}{cc}
\mu_1 & \mu_3   \\
\mu_4 & \mu_2  
\end{array}
\Big|1-\lambda).
\label{fu}
\end{eqnarray}
Note that $F_{\mu_{31}\mu_{23}}^{L}$ specifies the invertible 
fusion matrix in between $s$- and $t$-channel LFT 
conformal blocks. 
When one of four primary fields gets degenerate then 
${\cal B}$ reduces to a hypergeometric function such that 
\eqref{fu} reads 
\begin{eqnarray*}
 &&{}_2F_1 (a,b,c;\lambda)=\dfrac{\Gamma(c)\Gamma(c-a-b)}{\Gamma(c-a)\Gamma(c-b)}{}_2F_1 (a,b,a+b-c+1;1-\lambda) \\
&&+
\dfrac{\Gamma(c)\Gamma(a+b-c)}{\Gamma(a)\Gamma(b)}(1-z)^{c-a-b}{}_2F_1 (c-a , c-b, c-a-b+1; 1-\lambda)
\end{eqnarray*}
as having been indicated by 
connection coefficients listed in (2.2).

Notice that (I)-(VI) collected in Sec. 3.2 is somewhat unexpected and independent of (2.2). 
One can just hardly regard 
(2.2) as any evidence showing triality (or $S$-duality) 
as AGT conjecture relates Liouville conformal blocks to 
Nekrasov's instanton partition functions. This is because the latter never splits 
into two compatible (and linearly independent) components. 
On the other hand, given certain Riemann scheme dictated by 
four flavor masses, the permutation between 
six Kummer's solutions upon imposing triality ultimately amounts to another geometric 
realization of ${\mathbf S}_3$ outer automorphism of 
${\mathbb P}^1_{\ast}$. This picture will never 
be obtained by merely applying the crossing symmetry though.

\subsection*{(II) Geometric realization 
of triality} 

In order to make clearer the above statement, let us first 
write down the 
following expansion: 
\begin{eqnarray*}
\Psi(\lambda)={}_2F_1 \big(
\frac{a}{\hbar},\frac{b}{\hbar},\frac{c}{\hbar};\lambda \big)=
\exp -\Big(\hbar^{-1} W_0 + W_1 + \hbar W_2 +\cdots \Big).
\end{eqnarray*}
The zeroth-order term $W_0$ is obtained by means of the WKB 
approximation 
\begin{eqnarray*} 
\exp(-\hbar^{-1} W_0)= \exp\big(i\hbar^{-1}\int^{\lambda} dz \sqrt{T(z)}\big)
\end{eqnarray*}
w.r.t. the $Q$-form 
$\Big( \hbar^2\partial^2_{\lambda} + T({\lambda})\Big)\Psi({\lambda})=0$ 
of an usual hypergeometric equation%
\footnote{$\hbar$ is restored momentarily for computational 
convenience during applying the WKB method.}. It has been 
found in [10] that 
\begin{eqnarray*} 
&&W_0= \int^{\lambda} dz \sqrt{\frac{-a_1^2 z(1-z) + 
a_2^2(1-z) +
a_3^2 z}{z^2(1-z)^2}},\\
&& a= -a_1+a_2-a_3, ~~~~~b=a_1+a_2+a_3,~~~~~ 
c=2a_2.
\end{eqnarray*}
In addition, 
the curve 
\begin{eqnarray*} 
{\cal P}:~ y^2=T(z)= \frac{-a_1^2 z(1-z) + 
a_2^2(1-z) +
a_3^2 z}{z^2(1-z)^2}
\end{eqnarray*}
turns out to be dual to ${\mathbb P}^1_{\ast}$ 
[10, 28] since it stands for the real two-dimensional portion of 
M5-branes occupying ${\mathbb R}^{1,3} \times 
{\cal P}$ which arise from the uplift of 
a type IIA brane configuration (hence $\epsilon_1 + \epsilon_2=0$) engineering 
${\cal T}_{0,3}(A_1)$ theory of 
four free hypermultiplets. Therefore, choosing one Riemann scheme means fixing all parameters of 
${\cal P}$ (pants diagram), and interchanging 
solutions to its corresponding $Q$-form is 
equivalent to 
manifesting the outer automorphism ${\mathbf S}_3$ of ${\cal P}$. 

In summary, ${\mathbf S}_3$ originally stemming from 
the outer automorphism of $spin(8)$ 
(as seen from its Dynkin diagram) at the low-energy regime has been 
once interpreted as 
${\mathbf S}_4/{\mathbb Z}_2 \times {\mathbb Z}_2$ built 
on a four-punctured sphere via AGT dictionary (1.3). 
Through studying (1.1) and (1.2) under the 
hypergeometric nature of this sort of $SU(2)$ 
instanton partition functions, we see the revival of 
${\mathbf S}_3$ 
in a fashion naturally encoded in ${\cal P}$ 
dual to ${\mathbb P}^1_{\ast}$. 
This presentation dose provide a new geometric though simple viewpoint of (1.1) and (1.2) spelt out by Seiberg and 
Witten.

\section{Summary} 

Let us briefly summarize our 
main results. The 
geometric origin of triality stemming from 
the outer automorphism group 
${\mathbf S}_3$ of 
the quantum flavor symmetry $Spin(8)$ has long 
been pursuit. In Vafa's F-theory setup, 
a $D_4$-type singularity on an elliptically 
fibered K3 can be used to engineer 
an $\N=2$ $A_1$-type $N_f=4$ SCFT 
due to an arbitrary string coupling. 
While Vafa's picture compactified down to IIB theory stresses 
a geometric realization of $u$-plane parameterizing 
Coulomb branch, triality, namely \eqref{ma1} and 
\eqref{ma}, connecting physically 
equivalent theories seems 
not immediately visible. This is 
because now one is 
confined nearby a slightly deformed 
$D_4$-type singularity whereas 
in addition to bare mass parameters (positions of D7-branes 
located on $u$-plane) triality involves further an asymptotic piece of information, 
say, $\tau_0$ at $u\to \infty$. 
This problem of $\tau_0$ can be once 
remedied if one notices a 
bijection between the 
``fundamental" domain 
${\mathbb H}/{\Gamma(2)}$ of $\tau_0$ and 
moduli space of four 
marked points on a Riemann sphere 
by means of the celebrated $\lambda$-function \eqref{la}. 
The latter object denoted as $C_{0,4}$ emerges in 
Gaiotto's revolutionary description of an 
$\N=2$ $SU(2)$ $N_f=4$ SCFT. Instead, how to encode 
mass transformation rules into $C_{0,4}$ 
now turns out to be invisible.

What comes to one's rescue is AGT conjecture which 
states precisely \eqref{ttt}. 
Equipped with it, \eqref{ma1} and \eqref{ma} 
performed onto 
bare masses contained in 
$Z_{\text{inst}}$ as well as $\tau_0$ are then 
translated into interchanging six 
hypergeometric functions belonging to three 
regular singularities under 
certain Riemann scheme, provided one primary insertion 
of the four-point spherical conformal block gets degenerate. 
These arguments do provide another insight 
into capturing triality geometrically, e.g. 
permutation around vertices of a Schwarz triangle. 
Note that solutions in \eqref{51} are not equal to one another 
echoes the fact that ${\cal B}[C_{0,4}]$ along is basically 
not $S$-duality invariant or Nekrasov's partition function 
on ${\mathbb R}^4$ transforms nontrivially 
under $S$-duality as stressed in \cite{Alday:2009fs}.

\section*{Acknowledgements} 
I thank two Japanese mathematicians 
Masaaki Yoshida and Keiji Matsumoto for 
their e-mail correspondence and providing me 
with many valuable references. I am 
grateful to 
organizers of the workshop ``Recent Advances in Gauge Theories and CFTs" held at YITP Kyoto. I am also indebted to 
Toru Eguchi, Kazuhiro Sakai and Yuji Tachikawa for 
encouragement and helpful discussions. I am supported in part by the postdoctoral program at RIKEN.


\begin{thebibliography}{999}
\parskip=-2pt

\bibitem{Seiberg:1994aj}
  N.~Seiberg and E.~Witten,
  Nucl.\ Phys.\  B {\bf 431} (1994) 484
  [arXiv:hep-th/9408099].
  
\bibitem{Vafa:1996xn}
  C.~Vafa,
  Nucl.\ Phys.\  B {\bf 469} (1996) 403
  [arXiv:hep-th/9602022].
\bibitem{Sen:1996vd}
  A.~Sen,
  Nucl.\ Phys.\  B {\bf 475} (1996) 562
  [arXiv:hep-th/9605150].


    
\bibitem{Gaiotto:2009we}
  D.~Gaiotto,
  arXiv:0904.2715 [hep-th]. 
 
\bibitem{Alday:2009aq}
  L.~F.~Alday, D.~Gaiotto and Y.~Tachikawa,
  Lett.\ Math.\ Phys.\  {\bf 91} (2010) 167
  [arXiv:0906.3219 [hep-th]].

  





\bibitem{Alday:2010vg}
  L.~F.~Alday and Y.~Tachikawa,
  Lett.\ Math.\ Phys.\  {\bf 94} (2010) 87
  [arXiv:1005.4469 [hep-th]].







\bibitem{Teschner:2010je}
  J.~Teschner,
  arXiv:1005.2846 [hep-th].





\bibitem{Awata:2010yy}
  H.~Awata and Y.~Yamada,
  arXiv:1004.5122 [hep-th].






\bibitem{Morozov:2010cq}
  A.~Morozov and S.~Shakirov,
  JHEP {\bf 1008} (2010) 066
  [arXiv:1004.2917 [hep-th]].


\bibitem{Kozcaz:2010af}
  C.~Kozcaz, S.~Pasquetti and N.~Wyllard,
  JHEP {\bf 1008} (2010) 042
  [arXiv:1004.2025 [hep-th]].


\bibitem{Hadasz:2010xp}
  L.~Hadasz, Z.~Jaskolski and P.~Suchanek,
  JHEP {\bf 1006} (2010) 046
  [arXiv:1004.1841 [hep-th]].



\bibitem{Mironov:2010ym}
  A.~Mironov, A.~Morozov and A.~Morozov,
  arXiv:1003.5752 [hep-th].




\bibitem{Itoyama:2010ki}
  H.~Itoyama and T.~Oota,
  Nucl.\ Phys.\  B {\bf 838} (2010) 298
  [arXiv:1003.2929 [hep-th]].

\bibitem{Passerini:2010pr}
  F.~Passerini,
  JHEP {\bf 1003} (2010) 125
  [arXiv:1003.1151 [hep-th]].

\bibitem{Drukker:2010jp}
  N.~Drukker, D.~Gaiotto and J.~Gomis,
  arXiv:1003.1112 [hep-th].



\bibitem{Nekrasov:2010ka}
  N.~Nekrasov and E.~Witten,
  JHEP {\bf 1009} (2010) 092
  [arXiv:1002.0888 [hep-th]].


\bibitem{Popolitov:2010bz}
  A.~Popolitov,
  arXiv:1001.1407 [hep-th].

\bibitem{Chen:2010jga}
  B.~Chen, E.~O.~Colgain, J.~B.~Wu and H.~Yavartanoo,
  JHEP {\bf 1004} (2010) 078
  [arXiv:1001.0906 [hep-th]].

\bibitem{Mironov:2010zs}
  A.~Mironov, A.~Morozov and S.~Shakirov,
  Int.\ J.\ Mod.\ Phys.\  A {\bf 25} (2010) 3173
  [arXiv:1001.0563 [hep-th]].
\bibitem{Shakirov:2009nx}
  S.~Shakirov,
  arXiv:0912.5520 [hep-th].

\bibitem{Sulkowski:2009ne}
  P.~Sulkowski,
  JHEP {\bf 1004} (2010) 063
  [arXiv:0912.5476 [hep-th]].



\bibitem{Taki:2009zd}
  M.~Taki,
  arXiv:0912.4789 [hep-th].

\bibitem{Fujita:2009gf}
  M.~Fujita, Y.~Hatsuda and T.~S.~Tai,
  JHEP {\bf 1003} (2010) 046
  [arXiv:0912.2988 [hep-th]].

\bibitem{Alba:2009ya}
  V.~Alba and A.~Morozov,
  Nucl.\ Phys.\  B {\bf 840} (2010) 441
  [arXiv:0912.2535 [hep-th]].


\bibitem{Giribet:2009hm}
  G.~Giribet,
    JHEP {\bf 1001} (2010) 097 
  arXiv:0912.1930 [hep-th].


\bibitem{Fateev:2009aw}
  V.~A.~Fateev and A.~V.~Litvinov,
  JHEP {\bf 1002} (2010) 014
  [arXiv:0912.0504 [hep-th]].
\bibitem{m1}
A.~Mironov, A.~Morozov and S.~Shakirov,
  JHEP {\bf 1002} (2010) 030
  [arXiv:0911.5721 [hep-th]].



\bibitem{Schiappa:2009cc}
  R.~Schiappa and N.~Wyllard,
  arXiv:0911.5337 [hep-th].


\bibitem{Eguchi:2009gf}
  T.~Eguchi and K.~Maruyoshi,
  JHEP {\bf 1002} (2010) 022
  [arXiv:0911.4797 [hep-th]].

\bibitem{Kanno:2009ga}
  S.~Kanno, Y.~Matsuo, S.~Shiba and Y.~Tachikawa,
  Phys.\ Rev.\  D {\bf 81} (2010) 046004
  [arXiv:0911.4787 [hep-th]].

\bibitem{Itoyama:2009sc}
  H.~Itoyama, K.~Maruyoshi and T.~Oota,
  Prog.\ Theor.\ Phys.\  {\bf 123} (2010) 957
  [arXiv:0911.4244 [hep-th]].


\bibitem{Mironov}
  A.~Mironov and A.~Morozov,
  J.\ Phys.\ A  {\bf 43} (2010) 195401
  [arXiv:0911.2396 [hep-th]].
  
  
\bibitem{Hadasz:2009db}
  L.~Hadasz, Z.~Jaskolski and P.~Suchanek,
  JHEP {\bf 1001} (2010) 063
  [arXiv:0911.2353 [hep-th]].




\bibitem{Gaiotto:2009fs}
  D.~Gaiotto,
  arXiv:0911.1316 [hep-th].

\bibitem{Alba:2009fp}
  V.~Alba and A.~Morozov,
  arXiv:0911.0363 [hep-th].
\bibitem{m2}
A.~Mironov and A.~Morozov,
  JHEP {\bf 1004} (2010) 040
  [arXiv:0910.5670 [hep-th]].




\bibitem{Awata:2009ur}
  H.~Awata and Y.~Yamada,
  JHEP {\bf 1001} (2010) 125
  [arXiv:0910.4431 [hep-th]].




\bibitem{Gadde:2009kb}
  A.~Gadde, E.~Pomoni, L.~Rastelli and S.~S.~Razamat,
  JHEP {\bf 1003} (2010) 032
  [arXiv:0910.2225 [hep-th]].

\bibitem{Alday:2009qq}
  L.~F.~Alday, F.~Benini and Y.~Tachikawa,
  Phys.\ Rev.\ Lett.\  {\bf 105} (2010) 141601
  [arXiv:0909.4776 [hep-th]].





\bibitem{Mironov:2009qn}
  A.~Mironov and A.~Morozov,
  Phys.\ Lett.\  B {\bf 682} (2009) 118
  [arXiv:0909.3531 [hep-th]].

\bibitem{Poghossian:2009mk}
  R.~Poghossian,
  JHEP {\bf 0912} (2009) 038
  [arXiv:0909.3412 [hep-th]].

\bibitem{Mars}
  A.~Marshakov, A.~Mironov and A.~Morozov,
  JHEP {\bf 0911} (2009) 048
  [arXiv:0909.3338 [hep-th]].

\bibitem{Dijkgraaf:2009pc}
  R.~Dijkgraaf and C.~Vafa,
  arXiv:0909.2453 [hep-th].
\bibitem{m3}
A.~Marshakov, A.~Mironov and A.~Morozov,
  Phys.\ Lett.\  B {\bf 682} (2009) 125 
  [arXiv:0909.2052 [hep-th]].

\bibitem{Drukker:2009id}
  N.~Drukker, J.~Gomis, T.~Okuda and J.~Teschner,
  JHEP {\bf 1002} (2010) 057
  [arXiv:0909.1105 [hep-th]].

\bibitem{Alday:2009fs}
  L.~F.~Alday, D.~Gaiotto, S.~Gukov, Y.~Tachikawa and H.~Verlinde,
  JHEP {\bf 1001} (2010) 113
  [arXiv:0909.0945 [hep-th]].



\bibitem{Nanopoulos:2009au}
  D.~V.~Nanopoulos and D.~Xie,
  Phys.\ Rev.\  D {\bf 80} (2009) 105015
  [arXiv:0908.4409 [hep-th]].

\bibitem{Nekrasov:2009rc}
  N.~A.~Nekrasov and S.~L.~Shatashvili,
  arXiv:0908.4052 [hep-th].


\bibitem{Mir}
  A.~Mironov and A.~Morozov,
  Nucl.\ Phys.\  B {\bf 825} (2010) 1
  [arXiv:0908.2569 [hep-th]]. 
\bibitem{Mir2}
A.~Mironov and A.~Morozov,
  Phys.\ Lett.\  B {\bf 680} (2009) 188
  [arXiv:0908.2190 [hep-th]].

\bibitem{Mironov:2009dr}
  A.~Mironov, S.~Mironov, A.~Morozov and A.~Morozov,
  arXiv:0908.2064 [hep-th].

\bibitem{Gaiotto:2009ma}
  D.~Gaiotto,
  arXiv:0908.0307 [hep-th].

\bibitem{Marshakov:2009gs}
A.~Marshakov, A.~Mironov and A.~Morozov,
  Theor.\ Math.\ Phys.\  {\bf 164} (2010) 831 
  [arXiv:0907.3946 [hep-th]].

\bibitem{Maruyoshi:2009uk}
  K.~Maruyoshi, M.~Taki, S.~Terashima and F.~Yagi,
  JHEP {\bf 0909} (2009) 086
  [arXiv:0907.2625 [hep-th]].

\bibitem{Drukker:2009tz}
  N.~Drukker, D.~R.~Morrison and T.~Okuda,
  JHEP {\bf 0909} (2009) 031
  [arXiv:0907.2593 [hep-th]].

\bibitem{Wyllard:2009hg}
  N.~Wyllard,
  JHEP {\bf 0911} (2009) 002
  [arXiv:0907.2189 [hep-th]].

\bibitem{Nanopoulos:2009xe}
  D.~Nanopoulos and D.~Xie,
  JHEP {\bf 0908} (2009) 108
  [arXiv:0907.1651 [hep-th]].
  
  
\bibitem{Eguchi}
  T.~Eguchi and K.~Sakai,
  JHEP {\bf 0205} (2002) 058
  [arXiv:hep-th/0203025], Adv.\ Theor.\ Math.\ Phys.\  {\bf 7} (2004) 419
  [arXiv:hep-th/0211213].
\bibitem{MY}
  M.~Yoshida, 
  ``Hyper Geometric Functions, My Love: Modular Interpretations of Configuration Spaces (Aspects of Mathematics),"
Friedrick Vieweg and Son (1997/10) 
  
\bibitem{Zamolodchikov:1995aa}
  A.~B.~Zamolodchikov and A.~B.~Zamolodchikov,
  Nucl.\ Phys.\  B {\bf 477} (1996) 577
  [arXiv:hep-th/9506136].

\bibitem{BPZ}
A. A. Belavin, A. M. Polyakov, and A. B. Zamolodchikov, 
``Infinite conformal symmetry in
two-dimensional quantum field theory," 
Nucl.\ Phys.\  B {\bf 241} (1984) 333. 

\bibitem{Fateev:2009me}
  V.~A.~Fateev, A.~V.~Litvinov, A.~Neveu and E.~Onofri,
  J.\ Phys.\ A  {\bf 42} (2009) 304011
  [arXiv:0902.1331 [hep-th]].
\bibitem{ka}
J. Kaneko, ``q-Selberg integrals and Macdonald polynomials," Ann. Sci. ´ Ecole
Norm. Sup. {\bf 29} (1996) 583. 




\bibitem{iwa}
Mathematical Society of Japan, 
``Iwanami Suugaku Jiten." 4th Japanese
ed., Iwanami Shoten, 2007. 
\end{thebibliography}
\end{document}